# Centralized Modularity of *N*-Linked Glycosylation Pathways in Mammalian Cells

**Pan-Jun Kim[1], Dong-Yup Lee[2,3,*], Hawoong Jeong[4,5,*]**

[1]Institute for Genomic Biology, University of Illinois at Urbana-Champaign, Urbana, Illinois, United States of America, [2]Department of Chemical and Biomolecular Engineering, National University of Singapore, Singapore, [3]Bioprocessing Technology Institute, Agency for Science, Technology and Research (A*STAR), Singapore, [4]Institute for the BioCentury, KAIST, Daejeon, South Korea, [5]Department of Physics, KAIST, Daejeon, South Korea

* E-mail: cheld@nus.edu.sg (DYL); hjeong@kaist.edu (HJ)

**Glycosylation is a highly complex process to produce a diverse repertoire of cellular glycans that are attached to proteins and lipids. Glycans are involved in fundamental biological processes, including protein folding and clearance, cell proliferation and apoptosis, development, immune responses, and pathogenesis. One of the major types of glycans, *N*-linked glycans, is formed by sequential attachments of monosaccharides to proteins by a limited number of enzymes. Many of these enzymes can accept multiple *N*-linked glycans as substrates, thereby generating a large number of glycan intermediates and their intermingled pathways. Motivated by the quantitative methods developed in complex network research, we investigated the large-scale organization of such *N*-linked glycosylation pathways in mammalian cells. The *N*-linked glycosylation pathways are extremely modular, and are composed of cohesive topological modules that directly branch from a common upstream pathway of glycan synthesis. This unique structural property allows the glycan production between modules to be controlled by the upstream region. Although the enzymes act on multiple glycan substrates, indicating cross-talk between modules, the impact of the cross-talk on the module-specific enhancement of glycan synthesis may be confined within a moderate range by transcription-level control. The findings of the present study provide experimentally-testable predictions for glycosylation processes, and may be applicable to therapeutic glycoprotein engineering.**

## Introduction

Carbohydrates are a basic cell constituent, and are one of the most abundant and diverse biopolymers in nature [1]. Complex carbohydrates have recently become widely recognized as more than just a metabolic energy source [2–6]. For example, the cell surface contains a layer of complex carbohydrates involved in signalling roles that are indispensable to multicellular organisms [2,7]. Glycosylation, the attachment of glycans (oligosaccharides) to proteins or lipids, is a ubiquitous post-translational modification that generates an extensive functional capability from a limited set of genes [8–10]. In contrast to gene and protein sequences, the glycosylated glycan sequences are not arranged in a simple linear chain [5]. Several monosaccharides can be placed simultaneously on a particular monosaccharide, forming branched structures that provide enormous glycan structural diversity.

Vertebrates, and especially mammals, have evolved a unique glycan repertoire which is structurally distinct from that of nonvertebrate organisms [2,8–10]. Mammalian cells are used as host cell systems for the production of many recombinant glycoproteins; these systems can synthesize properly folded proteins with glycans resembling those in human bodies [11,12]. *N*-Linked and *O*-linked glycans are the major contributors to the structure and function of mammalian secretory glycoproteins. *N*-Linked glycans are attached to asparagine residues of proteins, located within the Asn-X-Ser/Thr motif of amino acids, where X can be any amino acid except proline.

*N*-Linked glycosylation occurs co-translationally in the endoplasmic reticulum (ER) compartment. The addition of an oligosaccharide to the peptide at an early stage of glycoprotein synthesis allows the glycan to participate in the folding and quality control of a newly synthesized protein [13]. Upon successful folding of the protein and the trimming of some residues in the glycan, the glycoprotein migrates into the Golgi apparatus. Processing in the Golgi involves the removal of mannose groups and the addition of various monosaccharides to the growing glycan. The removal of the mannose groups is driven by mannosidases, and the addition of different monosaccharides is facilitated by specific glycosyltransferases. Thus, *N*-linked glycosylation pathways comprise consecutive enzymatic steps that rely on the glycan structures produced by the previous enzyme to produce the substrate for the next enzyme. The pathways formed in this process diverge when a glycan is a substrate for multiple enzymes, or converge when multiple glycan substrates all lead to the same product. Many glycan intermediates at different loci along the pathways, not necessarily glycans at the termini, can be secreted out of the Golgi to the targeted sites where they perform biological functions, such as mediating cell growth and development, cell-cell communication, immune recognition/response, and molecular homeostasis [2,7–10,14].

Recent advances in understanding the generic properties of complex networks, including various biological, technological, and social networks [15,16], allow for a quantitative examination of the organization of *N*-linked glycosylation pathways. This development in network research has been driven largely by the availability of massive digital records and statistical methods that permit network data to be collected and analyzed on a scale far larger than previously possible. The emerging results in complex network research have led to the realization that, notwithstanding the importance of individual molecules, cellular phenotype is a contextual attribute of seamless and quantifiable network patterns among numerous constituents [17]. Despite the key role of glycosylation pathways in sustaining many biological functions, their large-scale properties have not yet been characterized from a complex network perspective. Understanding the global organization of complex networks will provide valuable and perhaps unique topological information, and may also lead to a better understanding of the dynamical and evolutionary processes of the networks, as demonstrated in several other biochemical systems, such as metabolic networks and protein-protein interaction networks [18–22]. Here, we explore whether the organization of glycosylation pathways can be elucidated from a complex network perspective, by investigating the structural and regulatory properties of *N*-linked glycosylation pathways in mammalian cells. Our findings don't only have the implications in the organizing principle of cellular glycosylation processes, but also in the glycoprotein engineering to be applicable for therapeutic purposes.

## Results

### Topological Properties and Modularity

We constructed *N*-linked glycan biosynthetic pathways by incorporating ten typical *N*-linked glycosylation enzymes in mammalian cells and their substrate specificities (Table 1; see also Materials and Methods). These enzymes can accept multiple *N*-linked glycans as substrates, and are thus capable of generating a large number of glycan intermediates. Construction was initiated from 9-mannose glycan, the common precursor of *N*-linked glycans in the Golgi, and followed by biosynthetic steps to produce mainly complex-type glycans (Figure 1), giving rise to a glycosylation network composed of 638 glycans and 1499 enzymatic reactions (Figure 2A).

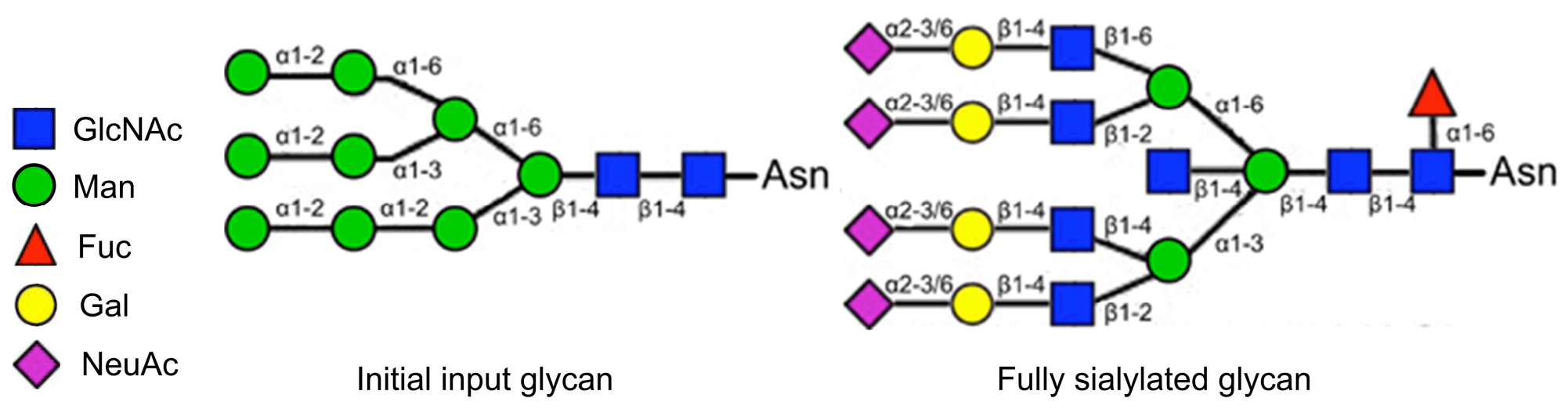

**Figure 1. Glycan structures considered in network construction.** 9-Mannose glycan enters the *N*-linked glycan biosynthetic pathways as the starting substrate (left), and can be processed into a fully sialylated complex-type glycan (right). Symbols indicate the sugar residues according to the CFG nomenclature [45]. GlcNAc: *N*-acetylglucosamine; Man: mannose; Fuc: fucose; Gal: galactose; NeuAc: sialic acid.

**Central and peripheral regions.** The essentiality of a particular glycan in the glycosylation network was assessed by counting the number of all downstream substrates that could not be produced in the absence of the given glycan. Following the terms in complex network research, this might be analogous to evaluating the *avalanche size* of a network after perturbing a single vertex [23–25]. Figure 2B shows that for most glycans (95.8%) the absence of an individual glycan did not affect any glycan production or only hindered the production of fewer than three glycans. On the other hand, the impact of the removal of the few remaining glycans (4.2%) spread over a wide range, even up to the damage at the whole system level. These minor, but highly-impacting glycans tended to be located adjacent to each other, thereby occupying a single clustered region in the pathways. Therefore, the clustered region could be easily distinguished from the other parts of the network, and was termed the *central* region (Materials and Methods). The central region consisted of one connected component of glycans, including the initial input substrate, and the non-central or *peripheral* region was bound to and derived from this central region.

**Modular structure.** The spectral method developed for graph partitioning (Materials and Methods) revealed that the peripheral region comprised 21 tightly-knit subgraphs. These 21 subgraphs or modules are densely connected groups of glycans, with only sparser connections between groups. Therefore, the modules tend to be biosynthetically isolated from each other. This biosynthetically-modular property of the pathways originates from the substrate specificity of the enzymes considered here, as described below, rather than from simple differences in individual glycan structures that the conventional scheme for glycan classification [2] has been based on. Interestingly, each module in the peripheral region was generated from only a few roots, all of which belonged to the central region (Figures 2A and S1). In other words, *N*-linked glycosylation pathways organized their modular structure in a highly centralized manner; the central region with a small number of glycans proliferated directly into all 21 modules in the peripheral region, thereby forming a star-like structure. Indeed, direct connections between different peripheral modules were relatively sparse compared with those between the central and peripheral regions (Figures S1 and S2). Remarkably, the glycosylation network had unusually high modularity ($Q$ = 0.83) compared with other biological and non-biological networks [26], suggesting that glycosylation-specific evolutionary pressure was required for the development of such a unique network structure.

The number of glycans across modules was unevenly distributed, with the largest module containing 40-fold more glycans than the smallest module. The discrete jumps between module sizes in Figure 2C indicate that the size of each module was due to the complexity of the terminal glycan structures. Specifically, the more processed the terminal glycans were with *N*-acetylglucosamines (GlcNAc) following α1,3- and α1,6-linked mannoses, the greater the number of glycan species that developed in the module. This glycan enrichment pattern across modules comes from the inherent capability of carbohydrates to add branches [5] to the mannose residues, which exponentially diversifies the glycan structures.

**Enzymatic contribution.** To better understand the modular properties of the glycosylation network, we further investigated the enzymatic reactions involved in module formation. Reactions from the central to the peripheral region – entry reactions into the peripheral modules – were dominated by galactosylation (93.5% of the reactions; see also Figures 2A and S1). Galactosylation and concurrent sialylation also dominated the reactions (99.0%) occurring *within* the peripheral modules (Figure 2A), while the reactions *between* these modules mostly comprised GlcNAc addition (Figure S2). On the other hand, in the central region, all enzymes except galactosyltransferase (GalT) and sialyltransferase (SiaT) were involved in the reactions (Figure 2A).

These findings suggest the enzymatic mechanisms that are responsible for generating the unique modular structure of the glycosylation network, as highlighted by the role of GalT: GalT are generous in their substrate specificity, accepting any substrate with free GlcNAc on the mannose branches, and multiple products arise from the same substrate depending on the specific galactosylated residues. Once glycans are galactosylated, however, they inhibit the approach of many other enzymes (Table 1). Such tolerance in substrate specificity and product formation facilitates the development of redundant pathways within each module, whereas the inhibition of other enzyme activities keeps different modules separated. The effect of such inhibition for module differentiation was also observed from another enzyme, β-1,4-mannosyl-glycoprotein 4-β-*N*-acetylglucosaminyltransferase (GnTIII). GnTIII adds bisecting GlcNAc to its substrate, and the presence of bisecting GlcNAc inhibits the activity of many enzymes (Table 1). Therefore, the bisecting GlcNAc is thought to insulate relevant modules, as shown in Figure S2 where the junctions of different modules only contain glycans without bisecting GlcNAc. Accordingly, if we exclude the glycan syntheses catalyzed by GnTIII, then the network becomes slightly less modular ($Q$ = 0.72) as the well-insulated modules selectively disappear. We believe that this organizing principle of modular structures manifested by GalT as well as by GnTIII offers a useful guideline for the engineering of novel glycosyltransferases, as discussed below.

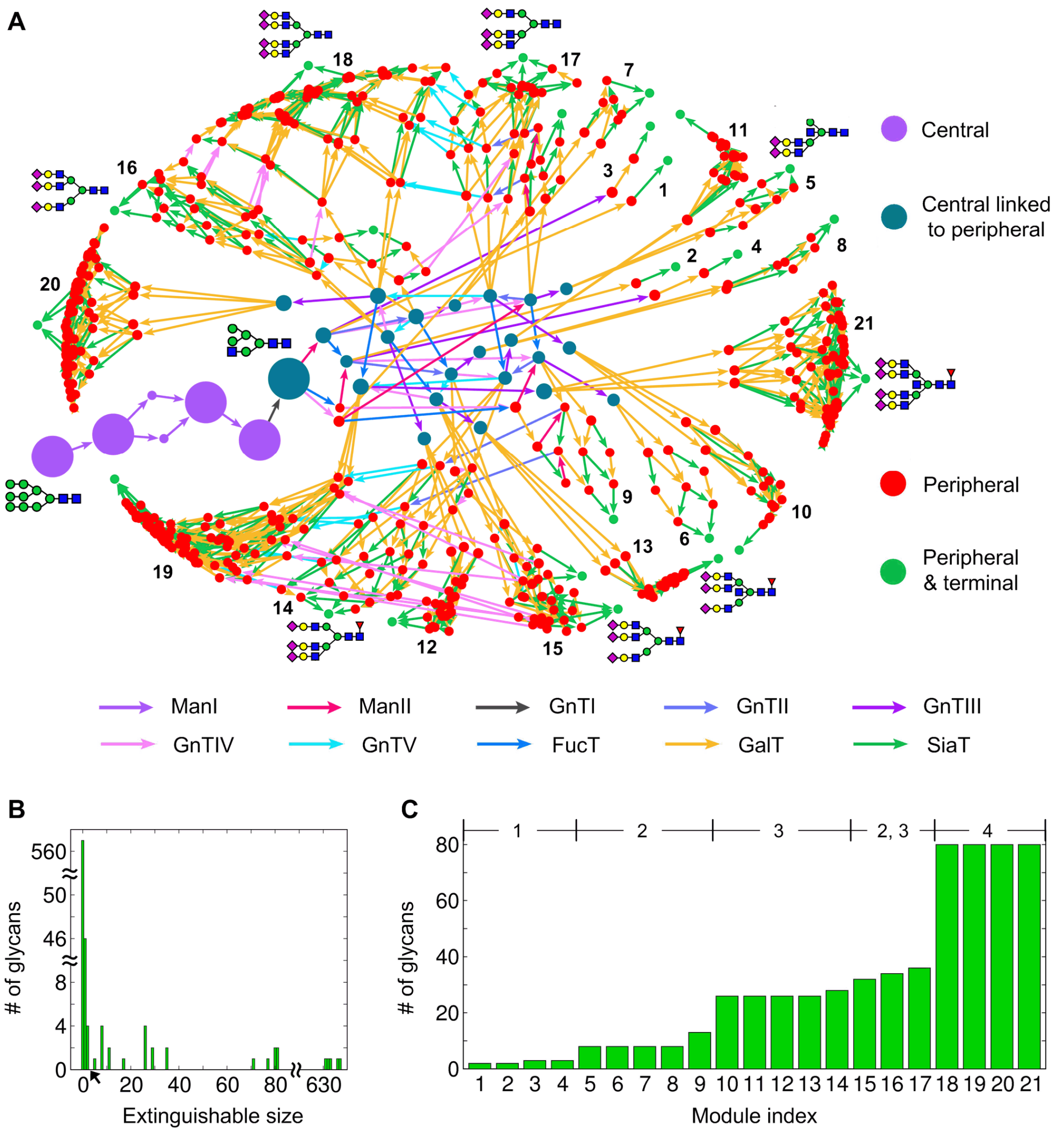

**Figure 2. Modular organization of *N*-linked glycosylation pathways.** (A) The global topology of the constructed *N*-linked glycosylation pathways. Circles stand for glycan species and arrows for enzymatic reactions from substrates to products. The size of each circle represents the impact on the pathways in the absence of the corresponding glycan. Circles and arrows are colored according to the positions in the pathways and the catalyzing enzymes, respectively. We depict the structures of some early and terminal glycans using the symbols in Figure 1. Glycan synthetic modules are labelled in the ascending order of the number of the participating glycans. (B) In the absence of each glycan, the number of extinguished downstream glycans is shown on the horizontal axis, and the vertical axis shows the number of such absent glycans leading to the same effect. The criterion of the central region is indicated by the arrow at the horizontal axis (Materials and Methods). (C) The number of glycans participating in each module. Module indices on the horizontal axis follow those appearing in (A). At the top in the horizontal direction, we show the number of GlcNAc on mannose branches that each terminal glycan has in a corresponding module.

---

## Regulatory Properties and Cross-Talk

The highly modularized, yet centralized organization of *N*-linked glycosylation pathways raises the question of how cells enhance or suppress the glycan production across modules against distinct physiological conditions. Within the same module, glycans are easily convertible to other glycans along densely connected pathways, whereas the conversion of glycans between different modules, which are only sparsely connected, is more difficult. Furthermore, glycans in the peripheral modules are surrounded by homogeneous enzymatic reactions (catalyzed mostly by GalT and SiaT in the Golgi), and are thus not as likely to be regulated but routed randomly along the pathways. Glycans along such unregulated routes are thought to be trapped for a long time in a particular module because there are few paths through which they can enter the other modules [27]; therefore, glycans delivered from the central region might continue to be processed inside the arrival modules until they are eventually secreted out of the Golgi. In this regard, the paths glycans take through the central region *ahead* of the peripheral modules likely play a critical role in the end-product formation.

Specific reactions in the central region may be manipulated by the transcriptional regulation of enzyme expression. Previous experiments demonstrated a correlation between glycan production and transcript expression of the corresponding enzymes. For example, the abundance of bisected glycoforms and of GnTIII transcript as well as that of fucosylated glycoforms and of glycoprotein 6-α-L fucosyltransferase (FucT) transcript is positively correlated across different mouse tissues [28–30]. The heterogeneous enzyme pools in the central region favor such specific transcriptional control. Glycosylation enzymes, however, are usually involved in multiple reactions; a change in the abundance of a single enzyme is likely to affect more than one reaction in the central region, and a number of modules derived from the affected reactions will also be affected. Therefore, it is important to assess specifically how to control these modules that share the common upstream enzymes to result in the cross-talk between the modules.

We considered combinations of up- and down-regulation of glycosylation enzymes that would unambiguously predict changes in glycan syntheses, and for each case, we determined which modules would enhance or suppress glycan production relative to their basal levels (Materials and Methods). Figure 3A shows one such result in which the down-regulation of GnTIII, α-1,3-mannosyl-glycoprotein 4-β-*N*-acetylglucosaminyltransferase (GnTIV), and FucT led to the enhancement of the 1st and 16th modules, but also to the suppression of the other modules. Minimizing cross-talk or unwanted enhancement of modules other than those specified requires an orchestrated regulation across enzymes. Under the regulation to minimize such cross-talk, Figures 3B and 3C show that each enhancement of three-quarters of the modules was accompanied by the unwanted enhancement of less than one-third of the modules, and the enhancement of the remaining modules could be at most accompanied by the unwanted enhancement of less than one-half of the modules. Consequently, although the cross-talk between modules is not negligible, the effect on glycan synthesis is confined within a moderate range, and probably further reduced by post-transcriptional regulation or by other combinations of enzyme regulation which were excluded here for clarity.

The explicit prediction of modules to be enhanced under given transcriptional regulation (Table S1) can be tested experimentally by measuring the change in the glycan production after genetic manipulation and identifying the relevant modules. For example, the production of glycans belonging to the 1st and 16th modules (Figure S1) is supposed to be increased after gene knockdown of GnTIII, GnTIV, and FucT, as indicated in Figure 3A. It should be noted that the glycan production here was quantified by the amount of flux into the glycan synthesis, rather than by the glycan abundance itself. Therefore, measuring only the abundance of secreted glycans and not the abundance of all the glycan intermediates will be more relevant in this case. Experimental validation of this prediction will allow us to design genetic regulation to enhance glycan synthesis in targeted modules. For example, if some modules contain desirable glycoproducts like biopharmaceuticals, then genetic regulation can be applied to enhance the glycan synthesis in these modules, and accordingly, to increase the production rate of the biopharmaceutical glycans. Such genetic regulation toward specific module enhancement might also be applied to reduce the heterogeneity of glycoforms and to improve the consistency of glycoprotein production [11,12].

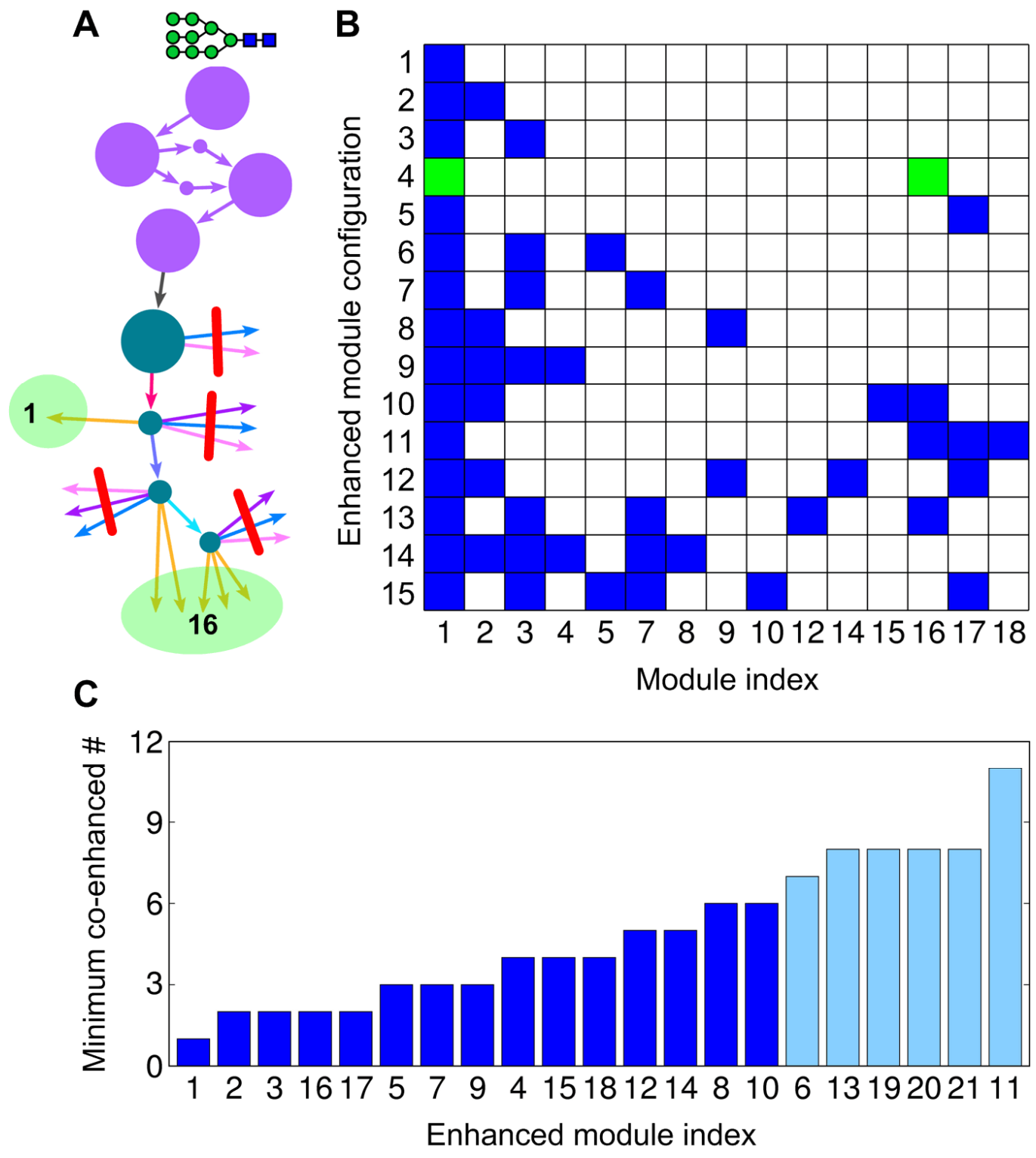

**Figure 3. Regulation of glycan synthesis.** (A) Enhancement of glycan production in specific modules is illustrated with a part of the pathways in Figure 2(A). The down-regulation of GnTIII, GnTIV, and FucT weakens the reactions crossed by red lines, thus strengthening the other reactions toward the 1st and 16th modules, as highlighted. (B) The modules to be enhanced or suppressed under combinations of enzyme up- and down-regulation. Here we show only the cases where six or fewer modules become enhanced, and exclude the indices of the modules that are never enhanced in these cases. Each row displays a unique pattern of module enhancement resulting from certain regulatory combinations, and is labelled as shown in Table S1. Enhanced modules are colored blue or green, and suppressed ones are indicated in white. The row including greens is for the case demonstrated in (A). (C) From the lists of modules to be enhanced together with a given module on the horizontal axis, we enumerated the minimum number of such co-enhanced modules as shown in the vertical axis. Blue is for less than or equal to 6 in the minimum number, obtained from the module enhancement pattern shown in (B).

## Discussion

The complexity and biological significance of protein glycosylation have long been underestimated, and now, in the post-genomic era, are at the forefront of scientific research. It is increasingly appreciated that biological systems exploit glycosylation in synthesizing cell-surface glycans to organize plasma membrane receptors and control the recruitment of intracellular signal transduction mediators. Hence, further knowledge of glycobiology will contribute to deciphering a myriad of biological phenomena. Clearly, a systems-level understanding of glycosylation processes will advance such scientific achievement. The *N*-linked glycosylation pathways comprise very distinct topological modules, all directly stemming from the common upstream pathway termed the central region. This central region might act as a 'control tower' of glycan production by redistributing glycan synthesis fluxes over the modules to adapt to different physiological conditions. Cross-activation or cross-talk between the modules, however, will restrict the fine-tuning level of the flux distribution. The topological properties of such *N*-linked glycosylation pathways were elucidated from a complex network viewpoint that further helps set the hypotheses on implicated functional and evolutionary properties.

The underlying mechanism of module development is clarified by the role of GalT, which accepts a wide range of substrates and makes multiple products to inhibit many other enzyme activities. The tolerance in glycan synthesis and the inhibition of other enzyme activities contribute to module formation and differentiation, respectively, while the latter is also observed similarly in the case of GnTIII. The significant influence of GalT in pathway formation provides a pattern for the design of novel glycosyltransferases to implant another module that does not severely disturb the pre-existing pathways. Such construction or evolution of a new module would not significantly hamper the functioning of the old modules, and are thus favorable both for engineering purposes and for evolution, which could be facilitated [31] by this module-level modification. Specifically, the sugar residues attached by these novel enzymes should not inhibit galactosylation and sialylation. On the other hand, the enzymes should not accept already-galactosylated substrates. If these two rules are satisfied, then the enzymes will synthesize glycans at the central region introducing a new module in the peripheral region. Interestingly, GnTIII satisfies both rules, and this might be one reason why GnTIII works properly in recombinant Chinese Hamster Ovary (CHO) cells although it is not present in wild-type CHO [32–35]. In addition, CHO cells transfected with GnTIII are utilized in industry for the production of antibodies that significantly improve antibody-dependent cellular cytotoxicity and treat neuroblastoma and non-Hodgkin's lymphoma [34,35].

More immediate applications for glycoprotein engineering might arise from the relationship between transcriptional regulation and glycan production, as described above. Orchestrated regulation of enzyme expression in the central region will allow glycan production to be enhanced in specific modules, while avoiding moderately the increased production of other unwanted modules. Possible deviations between the prediction and the empirical data may arise due to incompleteness in modelling or regulation at a post-transcriptional level of which the potential effects on glycosylation remain largely unknown. Further integration of poly-*N*-acetyllactosamine structures and many degradation mechanisms will dress up the pathways considered here, and the original pathways can be viewed as an organizational kernel [36] of which the main properties we expect to be still reflected in more complicated pathways. Various techniques used to study metabolic flux analysis are also expected to allow for in-depth analysis of glycosylation processes [6]. In conjunction with such mathematical modelling [6,37,38], the development of high-throughput experimental techniques for glycan and glyco-gene profiling [3–5,10] will further facilitate the systems analysis of glycosylation processes as successfully demonstrated in this study.

**Table 1. Characteristics of enzymes used in network construction.** Substrate for each enzyme should fulfill all the required conditions. For the specific description of glycosidic linkages, refer to glycan structures in Figure 1. ManI: mannosyl-oligosaccharide 1,2-α-mannosidase; ManII: mannosyl-oligosaccharide 1,3-1,6-α-mannosidase; GnTI: α-1,3-mannosyl-glycoprotein 2-β-*N*-acetylglucosaminyltransferase; GnTII: α-1,6-mannosyl-glycoprotein 2-β-*N*-acetylglucosaminyltransferase; GnTIII: β-1,4-mannosyl-glycoprotein 4-β-*N*-acetylglucosaminyltransferase; GnTIV: α-1,3-mannosyl-glycoprotein 4-β-*N*-acetylglucosaminyltransferase; GnTV: α-1,6-mannosyl-glycoprotein 6-β-*N*-acetylglucosaminyltransferase; FucT: glycoprotein 6-α-L fucosyltransferase; GalT: β-*N*-acetylglucosaminylglycopeptide β-1,4-galactosyltransferase; SiaT: β-galactoside α-2,3/6-sialyltransferase.

| Enzyme | Substrate requirement | Catalyzing reaction |
|---|---|---|
| **ManI** | Free α1,2-linked mannose | Ordered removal of free α1,2-linked mannose (Materials and Methods) |
| **ManII** | Free α1,3- or α1,6-linked mannose following α1,6-linked mannose<br>Free β1,2-linked *N*-acetylglucosamine following α1,3-linked mannose<br>No bisecting β1,4-linked *N*-acetylglucosamine | Removal of free α1,3- and α1,6-linked mannoses |
| **GnTI** | Unique structure of 5-mannose glycan just processed by ManI | Addition of β1,2-linked *N*-acetylglucosamine to α1,3-linked mannose following β1,4-linked mannose |
| **GnTII** | No extra mannose other than three in the core<br>No β1,2-linked *N*-acetylglucosamine following α1,6-linked mannose<br>Free β1,2-linked *N*-acetylglucosamine following α1,3-linked mannose<br>No bisecting β1,4-linked *N*-acetylglucosamine | Addition of β1,2-linked *N*-acetylglucosamine to α1,6-linked mannose |
| **GnTIII** | No bisecting β1,4-linked *N*-acetylglucosamine<br>β1,2-linked *N*-acetylglucosamine following α1,3-linked mannose<br>No β1,4-linked galactose | Addition of bisecting β1,4-linked *N*-acetylglucosamine |
| **GnTIV** | No β1,4-linked *N*-acetylglucosamine following α1,3-linked mannose<br>Free β1,2-linked *N*-acetylglucosamine following α1,3-linked mannose<br>No bisecting β1,4-linked *N*-acetylglucosamine | Addition of β1,4-linked *N*-acetylglucosamine to α1,3-linked mannose |
| **GnTV** | No β1,6-linked *N*-acetylglucosamine<br>Free β1,2-linked *N*-acetylglucosamine following α1,6-linked mannose<br>No bisecting β1,4-linked *N*-acetylglucosamine | Addition of β1,6-linked *N*-acetylglucosamine |
| **FucT** | No α1,6-linked fucose<br>β1,2-linked *N*-acetylglucosamine following α1,3-linked mannose<br>No bisecting β1,4-linked *N*-acetylglucosamine<br>No β1,4-linked galactose | Addition of α1,6-linked fucose |
| **GalT** | Free *N*-acetylglucosamine following either of α1,3- or α1,6-linked mannose | Addition of β1,4-linked galactose to free *N*-acetylglucosamine following either of α1,3- or α1,6-linked mannose |
| **SiaT** | Free β1,4-linked galactose | Addition of sialic acid to free β1,4-linked galactose |

## Materials and Methods

### Network Construction

*N*-linked glycosylation pathways were constructed by enumerating *N*-linked glycan structures commonly observed in mammalian cells [39], starting from the input substrate shown in Figure 1, which results from an oligosaccharide precursor in the ER with three glucose residues trimmed out. In our attempts to build consecutive enzymatic steps, we used ten enzymes constituting a large proportion of the mammalian *N*-linked glycosylation processes. The mannosidases (ManI and ManII) are exoglycosidases that remove mannose groups from *N*-linked glycans. The other eight enzymes are glycosyltransferases that catalyze the formation of glycosidic bonds. Five *N*-acetylglucosaminyltransferases (GnTI, GnTII, GnTIII, GnTIV, and GnTV) were considered for the addition of GlcNAc, and FucT, GalT, and SiaT for the addition of fucose, galactose, and sialic acid, respectively.

Based on previous *in vivo* observations, the removal of α1,2-linked mannoses by ManI was considered in the following order [40,41]: ER resident ManI removes free α1,2-linked mannose attached to α1,3-linked mannose in the initial input substrate and then Golgi resident ManI removes each of two remaining free α1,2-linked mannoses successively, making 6-mannose and then 5-mannose glycans. For the remaining enzymatic reactions, we applied the substrate specificity data shown in Table 1 obtained from publicly available literatures [37,42,43]. Except GnTI which uses only one substrate, the other enzymes could catalyze reactions that involve the same glycosidic linkage on a range of different substrates. Finally, by taking into account only the pathways to be terminated at glycans containing mannoses no more than three in the core residue, we integrated pathways to produce mainly complex-type glycans for clarity of analysis. The resulting pathways are represented by a directional graph in which the vertices stand for glycan species and the edges for glycan synthetic reactions with arrows pointing from substrates to products.

### Network Decomposition into Subunits

The essentiality of individual glycans in the pathways was investigated by perturbing the pathways through the removal of single glycans. For each removal, we calculated how many glycans could not be produced due to the complete absence of their substrate production. The removal for most of the glycans gave only negligible effects (smaller than the cut-off in Figure 2B), and accordingly, we grouped the remaining glycans, whose removal had large effects, into those in the central region together with the early glycans processed by ManI. Glycans in the central region were located adjacent to each other, forming a self-jointed subgraph and containing root vertices linked to the non-central or peripheral region. Different criteria for the central region did not affect the main results presented here as long as the cut-off was set between 2~8 (Figure 2B).

The peripheral region could be further partitioned by maximizing modularity $Q$ for directional graphs [44]:

$$Q = \frac{1}{m}\sum_{ij}\left[A_{ij} - \frac{k_i^{\mathrm{in}} k_j^{\mathrm{out}}}{m}\right]\delta_{c_i,c_j},$$

where $A_{ij}$ is 1 if there is an edge from vertex $j$ to vertex $i$ and, otherwise 0. $k_i^{\mathrm{in}}$ and $k_j^{\mathrm{out}}$ are the numbers of incoming and outgoing edges of the vertices, $m$ is the total number of edges in the graph, $\delta_{ij}$ is the Kronecker delta symbol, and $c_i$ is the label of the partition to which vertex $i$ is assigned. Search for the division of the graph into partitions $\{c_i\}$ maximizing $Q$ is known to be NP-complete, thus we used the spectral optimization method [44], which is both computationally efficient and practically acceptable in terms of partitioning results. For this purpose, we pre-assigned the central region a partition and recursively decomposed the peripheral region based on the spectral method ($Q$ = 0.83). The resulting partitions or modules in the peripheral region were labelled in ascending order of the number of constituent glycans. The results of such partitioning on the glycosylation pathways remained robust when an alternative method that was designed for bidirectional or undirected graphs was applied by ignoring the edge directions [26]. Automatic decomposition of all the pathways, including the central region, yielded only a slight increase in modularity ($\Delta Q$ = 0.02), and this result was excluded to prevent method-specific over-partitioning that does not convey any information of biological significance.

## Glycan Synthesis Regulation

To evaluate the effect of transcriptional regulation on glycan synthesis, each enzyme $i$ was assigned variable $E_i$ depending on its regulated state: $E_i = 1$ if up-regulated, $E_i = 0$ if neutral, and $E_i = -1$ if down-regulated. In addition, let $G_{ij}$ be 1 if enzyme $i$ is involved in synthesizing glycan $j$ and, otherwise 0. Likewise, $M_{jk}$ is 1 if glycan $j$ is located at the entry of module $k$ and, otherwise 0. To focus on unambiguous cases in the prediction of regulatory effects, we only considered the combinations of $E_i$s that satisfied the following rules simultaneously:

(1) $E_i \cdot E_{i'} \geq 0$ for every pair of $i$ and $i'$ satisfying $G_{ij} = G_{i'j} = 1$, when the given $j$ and $k$ satisfy $M_{jk} = 1$. Hence, the mixture of both up- and down-regulated enzymes to synthesize a particular glycan at the entry of any given module was excluded.

(2) $\left(\sum_i E_i G_{ij}\right) \cdot \left(\sum_i E_i G_{ij'}\right) \geq 0$ for every pair of $j$ and $j'$ satisfying $M_{jk} = M_{j'k} = 1$ with a given $k$. Hence, the mixture of both enhanced and suppressed glycan production at the entry of a particular module was excluded.

Furthermore, we kept $E_{\mathrm{ManI}} = E_{\mathrm{SiaT}} = 0$ and $E_{\mathrm{GalT}} \geq 0$ to avoid an otherwise global and unspecific impact on glycan synthesis across modules. Each module $k$ could be assigned

$$\Phi_k = H\left(\sum_{i,j} E_i G_{ij} M_{jk}\right)$$ where $H(x)$ = 1, 0, or –1 if $x > 0$, $x = 0$, or $x < 0$, respectively.

Although $\Phi_k$ can be 1, 0, or –1, these three numbers did not appear simultaneously for any combination of $E_i$s. For example, some particular combination of $E_i$s allowed $\Phi_k$s to take 1 and 0, but never observed was a combination allowing them to take all of 1, 0, and –1. Here we considered the cases where two of 1, 0, and –1 were taken by $\Phi_k$s for given $E_i$s. Because we were interested in the regulatory cases keeping a similar level of the influx of the starting substrate for which modules compete with each other, modules assigned $\Phi_k$ larger than the other were expected to have enhanced glycan production relative to their basal levels, and were otherwise suppressed. (Table S1). For example, modules assigned $\Phi_k = 1$ were regarded as enhanced while the others assigned $\Phi_k = 0$ were regarded as suppressed. One can easily prove that such a regulatory effect remains invariant to applying both $E_i \rightarrow -E_i$ and permutations of enhanced and suppressed modules. We also examined alternative regulatory models, such as explicitly considering the substrate competition between reactions, but the results did not differ much from the present results.

## Acknowledgements

The authors wish to express gratitude to Wonjun Park, Le Thi Phuong Thao, and Doo-Byoung Oh for fruitful discussions and for providing valuable information. This work was supported by KRCF and Korean Systems Biology Program Grant M10309020000-03B5002-00000 (HJ), and the Biomedical Research Council of A*STAR (Agency for Science, Technology, and Research), Singapore (DYL), and the IGB Postdoctoral Fellows Program (PJK).

## Supplementary Materials

**Figure S1.** Entry and terminal glycans of peripheral modules. For each module, the parent glycans in the central region and the corresponding reactions are also depicted. The bulk of each module is dominated by galactosylation and sialylation.

**Figure S2.** Reactions between glycans belonging to different modules.

**Table S1.** Lists of enhanced or suppressed modules under combinations of enzyme regulation. For the comparison with Figure 3(B), the most right column labels each regulatory outcome in which no more than six modules become enhanced.

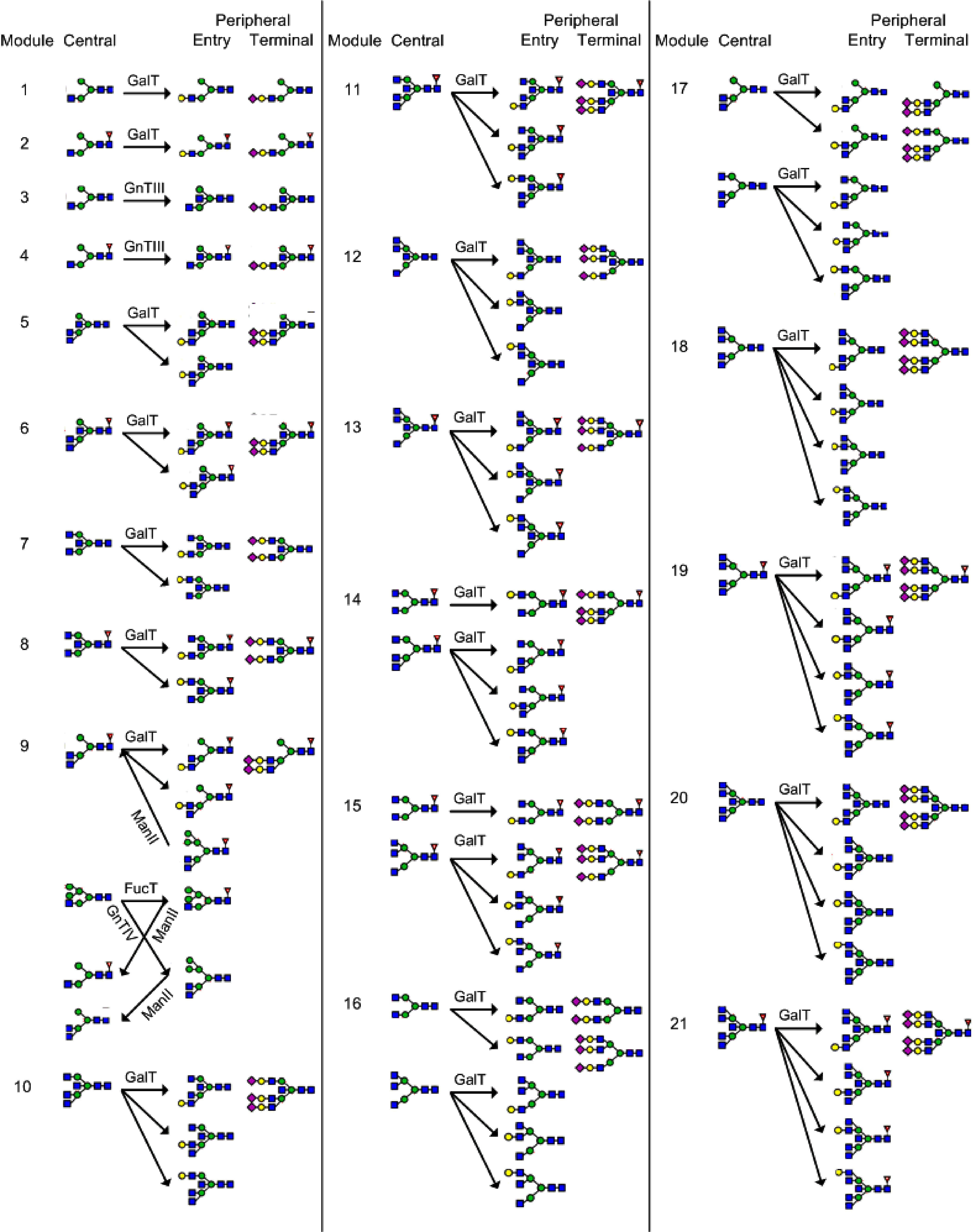

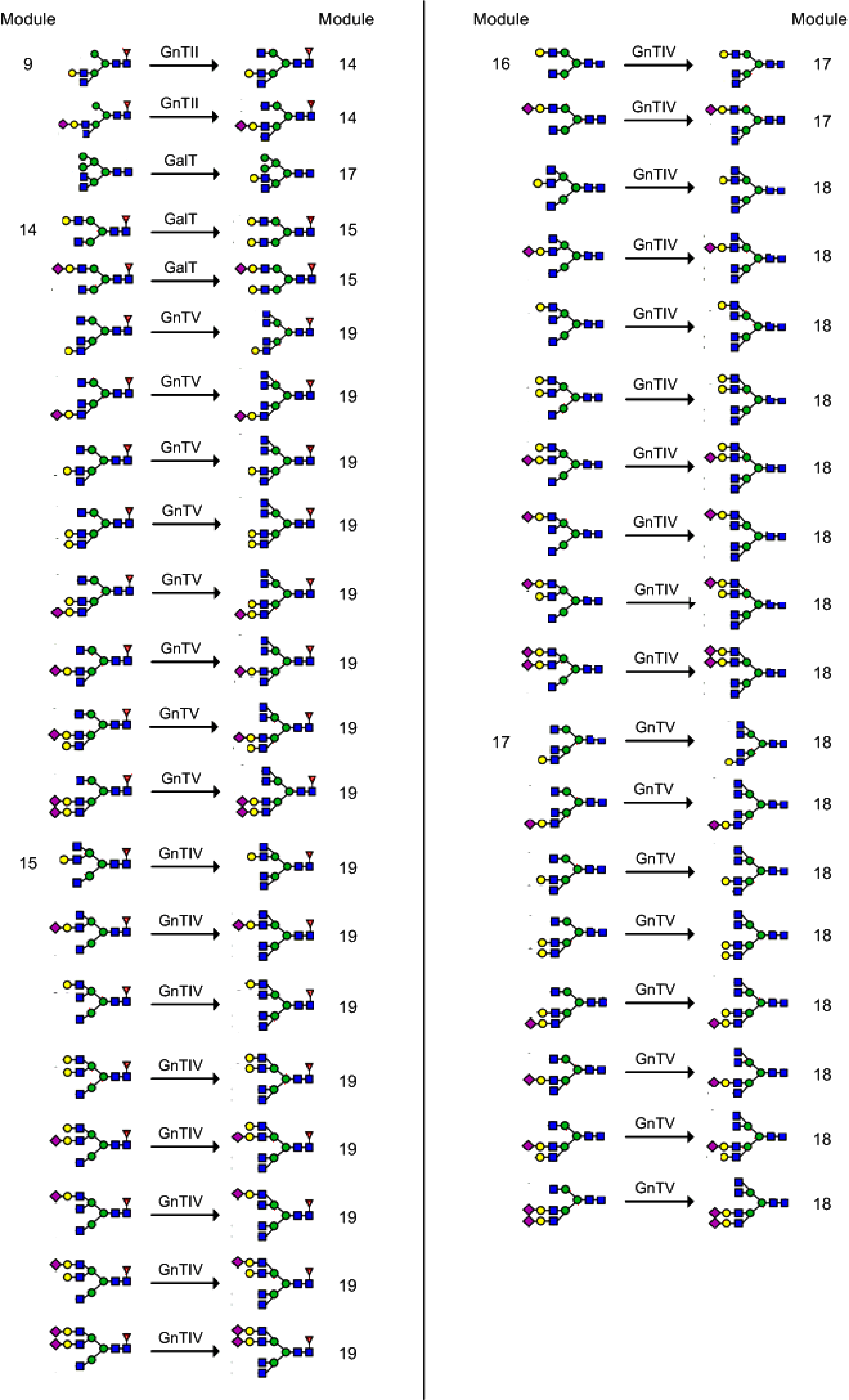

| Regulated Enzyme | | Affected Module | | |
|---|---|---|---|---|
| Up-regulated | Down-regulated | Enhanced | Suppressed | |
| | GnTII, GnTIII, FucT | 1 | 2~21 | 1 |
| | GnTII, GnTIII, GnTIV, FucT | 1 | 2~21 | |
| | GnTII, GnTIII, GnTIV, GnTV, FucT | 1 | 2~21 | |
| | GnTII, GnTIII, GnTV, FucT | 1 | 2~21 | |
| | GnTIII, GnTIV, GnTV, FucT | 1 | 2~21 | |
| | GnTII, GnTIII, GnTIV | 1, 2 | 3~21 | 2 |
| | GnTII, GnTIII, GnTIV, GnTV | 1, 2 | 3~21 | |
| | GnTIII, GnTIV, GnTV | 1, 2 | 3~21 | |
| | GnTII, GnTIV, FucT | 1, 3 | 2, 4~21 | 3 |
| | GnTII, GnTIV, GnTV, FucT | 1, 3 | 2, 4~21 | |
| | GnTIII, GnTIV, FucT | 1, 16 | 2~15, 17~21 | 4 |
| | GnTIII, GnTV, FucT | 1, 17 | 2~16, 18~21 | 5 |
| | GnTII, FucT | 1, 3, 5 | 2, 4, 6~21 | 6 |
| | GnTII, GnTV, FucT | 1, 3, 5 | 2, 4, 6~21 | |
| | GnTIV, GnTV, FucT | 1, 3, 7 | 2, 4~6, 8~21 | 7 |
| | GnTII, GnTIII | 1, 2, 9 | 3~8, 10~21 | 8 |
| | GnTII, GnTIII, GnTV | 1, 2, 9 | 3~8, 10~21 | |
| | GnTII, GnTIV | 1~4 | 5~21 | 9 |
| | GnTII, GnTIV, GnTV | 1~4 | 5~21 | |
| | GnTIII, GnTIV | 1, 2, 15, 16 | 3~14, 17~21 | 10 |
| | GnTIII, FucT | 1, 16~18 | 2~15, 19~21 | 11 |
| | GnTIII, GnTV | 1, 2, 9, 14, 17 | 3~8, 10~13, 15, 16, 18~21 | 12 |
| | GnTIV, FucT | 1, 3, 7, 12, 16 | 2, 4~6, 8~11, 13~15, 17~21 | 13 |
| | GnTIV, GnTV | 1~4, 7, 8 | 5, 6, 9~21 | 14 |
| | GnTV, FucT | 1, 3, 5, 7, 10, 17 | 2, 4, 6, 8, 9, 11~16, 18~21 | 15 |
| | GnTII | 1~6, 9 | 7, 8, 10~21 | |
| | GnTII, GnTV | 1~6, 9 | 7, 8, 10~21 | |
| GnTV | | 12, 13, 15, 16, 18~21 | 1~11, 14, 17 | |
| | GnTIII | 1, 2, 9, 14~19 | 3~8, 10~13, 20, 21 | |
| | GnTIV | 1~4, 7, 8, 12, 13, 15, 16 | 5, 6, 9~11, 14, 17~21 | |
| | FucT | 1, 3, 5, 7, 10, 12, 16~18, 20 | 2, 4, 6, 8, 9, 11, 13~15, 19, 21 | |
| FucT | | 2, 4, 6, 8, 9, 11, 13~15, 19, 21 | 1, 3, 5, 7, 10, 12, 16~18, 20 | |
| GnTIV | | 5, 6, 9~11, 14, 17~21 | 1~4, 7, 8, 12, 13, 15, 16 | |
| GnTIII | | 3~8, 10~13, 20, 21 | 1, 2, 9, 14~19 | |
| | GnTV | 1~11, 14, 17 | 12, 13, 15, 16, 18~21 | |
| GnTII | | 7, 8, 10~21 | 1~6, 9 | |
| GnTII, GnTV | | 7, 8, 10~21 | 1~6, 9 | |
| GnTV, FucT | | 2, 4, 6, 8, 9, 11~16, 18~21 | 1, 3, 5, 7, 10, 17 | |
| GnTIV, GnTV | | 5, 6, 9~21 | 1~4, 7, 8 | |
| GnTIV, FucT | | 2, 4~6, 8~11, 13~15, 17~21 | 1, 3, 7, 12, 16 | |
| GnTIII, GnTV | | 3~8, 10~13, 15, 16, 18~21 | 1, 2, 9, 14, 17 | |
| GnTIII, FucT | | 2~15, 19~21 | 1, 16~18 | |
| GnTIII, GnTIV | | 3~14, 17~21 | 1, 2, 15, 16 | |
| GnTIV, GnTV, FucT | | 2, 4~6, 8~21 | 1, 3, 7 | |
| GnTII, FucT | | 2, 4, 6~21 | 1, 3, 5 | |
| GnTII, GnTV, FucT | | 2, 4, 6~21 | 1, 3, 5 | |
| GnTII, GnTIII | | 3~8, 10~21 | 1, 2, 9 | |
| GnTII, GnTIII, GnTV | | 3~8, 10~21 | 1, 2, 9 | |
| GalT | | 1, 2, 5~21 | 3, 4 | |

| Regulated Enzyme | | Affected Module | |
|---|---|---|---|
| Up-regulated | Down-regulated | Enhanced | Suppressed |
| GnTII, GalT | | 1, 2, 5~21 | 3, 4 |
| GnTII, GnTIV, GalT | | 1, 2, 5~21 | 3, 4 |
| GnTII, GnTIV, GnTV, GalT | | 1, 2, 5~21 | 3, 4 |
| GnTII, GnTV, GalT | | 1, 2, 5~21 | 3, 4 |
| GnTIV, GalT | | 1, 2, 5~21 | 3, 4 |
| GnTIV, GnTV, GalT | | 1, 2, 5~21 | 3, 4 |
| GnTV, GalT | | 1, 2, 5~21 | 3, 4 |
| GnTIII, GnTV, FucT | | 2~16, 18~21 | 1, 17 |
| GnTIII, GnTIV, FucT | | 2~15, 17~21 | 1, 16 |
| GnTII, GnTIV, FucT | | 2, 4~21 | 1, 3 |
| GnTII, GnTIV, GnTV, FucT | | 2, 4~21 | 1, 3 |
| GnTII, GnTIII, GnTIV | | 3~21 | 1, 2 |
| GnTII, GnTIII, GnTIV, GnTV | | 3~21 | 1, 2 |
| GnTIII, GnTIV, GnTV | | 3~21 | 1, 2 |
| GnTII, GnTIV | | 5~21 | 1~4 |
| GnTII, GnTIV, GnTV | | 5~21 | 1~4 |
| FucT, GalT | | 1, 2, 4~21 | 3 |
| GnTII, FucT, GalT | | 1, 2, 4~21 | 3 |
| GnTII, GnTIV, FucT, GalT | | 1, 2, 4~21 | 3 |
| GnTII, GnTIV, GnTV, FucT, GalT | | 1, 2, 4~21 | 3 |
| GnTII, GnTV, FucT, GalT | | 1, 2, 4~21 | 3 |
| GnTIV, FucT, GalT | | 1, 2, 4~21 | 3 |
| GnTIV, GnTV, FucT, GalT | | 1, 2, 4~21 | 3 |
| GnTV, FucT, GalT | | 1, 2, 4~21 | 3 |
| GnTII, GnTIII, FucT | | 2~21 | 1 |
| GnTII, GnTIII, GnTIV, FucT | | 2~21 | 1 |
| GnTII, GnTIII, GnTIV, GnTV, FucT | | 2~21 | 1 |
| GnTII, GnTIII, GnTV, FucT | | 2~21 | 1 |
| GnTIII, GnTIV, GnTV, FucT | | 2~21 | 1 |